\documentclass[aps,pre,superscriptaddress,showkeys,floatfix,10pt]{revtex4-1}
\usepackage{graphicx}
\usepackage{dcolumn}
\usepackage{bm}
\usepackage{latexsym}
\usepackage{amssymb}
\usepackage{amsmath}
\begin{document}
\title{A generalized Michaelis-Menten equation in protein synthesis:\\ effects of mis-charged cognate tRNA and mis-reading of codon} 
\author{Annwesha Dutta}
\author{Debashish Chowdhury{\footnote{Corresponding author(E-mail: debch@iitk.ac.in)}}}

\affiliation{Department of Physics, Indian Institute of Technology,
Kanpur 208016, India.}
\begin{abstract}

The sequence of amino acid monomers in the primary structure of a 
protein is decided by the corresponding sequence of codons (triplets of 
nucleic acid monomers) on the template messenger RNA (mRNA). The 
polymerization of a protein, by incorporation of the successive amino 
acid monomers, is carried out by a molecular machine called ribosome. 
We develop a stochastic kinetic model that captures the possibilities of 
mis-reading of mRNA codon and prior mis-charging of a tRNA. 
By a combination of analytical and numerical methods we obtain 
the distribution of the times taken for incorporation of the successive amino 
acids in the growing protein in this mathematical model. The corresponding 
exact analytical expression for the average rate of elongation of a nascent 
protein is a `biologically motivated' generalization of the 
{\it Michaelis-Menten formula} for the average rate of enzymatic reactions. 
This generalized Michaelis-Menten-like formula (and the exact analytical 
expressions for a few other quantities) that we report here display the 
interplay of four different branched pathways corresponding to selection 
of four different types of tRNA.

\end{abstract}
\maketitle
\section{Introduction}

Enzymes are known to play crucial roles in almost all kinds of intracellular processes 
\cite{rittie08}. 
For the simplest enzymatic reaction studied theoretically by Michaelis and Menten 
more than a century ago \cite{MM13,boyde13} the rate of the formation of the 
product in bulk is given by the so-called Michaelis-Menten (MM) equation 
\cite{johnson13,michel13}. However, at extremely low population of an 
enzyme the time taken for each enzymatic cycle fluctuates from one cycle to 
another; the time taken in each round is often referred to as the turnover time. 
The distribution of the turnover time is the key statistical characteristics of reactions 
studied by single-molecule enzymology \cite{grima14}. Interestingly, in spite of 
the fluctuations in the turnover time, the mean turnover time for a large class of 
enzymatic reactions still follows the MM equation \cite{xie13a}. 
Over the last century various generalizations of the MM equation have emerged in 
several different contexts \cite{schnell03}. In this paper we present a generalization 
that emerges naturally in the context of biophysical chemistry of protein synthesis.

Proteins are polymers whose monomeric subunits are amino acids. 
The specific sequence of the amino acid species in the primary linear 
structure of a given protein is directed by the corresponding sequence 
of codons (triplets of nucleotide monomers) on the corresponding 
template messenger RNA (mRNA). The template-directed polymerization 
of a protein, called translation, is carried out by a molecular machine 
called ribosome \cite{spirin02,rodnina11,frank11,frank12,chowdhury13a,chowdhury13b} 
that consists of two loosely connected subunits designated as large and 
small.  Transfer RNA (tRNA) molecules play crucial roles in translation 
\cite{kim14}. When ``charged'' (amino-acylated) by a specific enzyme, 
called amino-acyl tRNA synthetase (aa-tRNA synthetase) 
\cite{ling09,reynolds10,yadavalli12}, 
one end of each species of these ``adapter'' molecules carries a 
specific amino acid. The amino acid brought in by a correctly charged 
cognate tRNA is also the correct one, as directed by the corresponding 
template; the other end of the same cognate tRNA molecule, referred to 
as anti-codon, matches perfectly, by complementary base pairing, with 
the codon on the template mRNA. 
In contrast, increasing degree of mismatch makes the tRNA near-cognate 
or non-cognate.

Most aa-tRNA synthetases employ editing mechanisms to ensure correct 
charging of the corresponding tRNA molecules. However, because of the 
intrinsic stochasticity of aminoacylation, and occasional failure of 
the editing mechanism of those aa-tRNA synthetase that are capable of 
correcting erroneous aminoacylation, a mis-charged tRNA may be produced 
\cite{ling09,yadavalli12}. 
Therefore, even when it turns out to be a cognate tRNA for a given codon, 
such a mis-charged tRNA compromises the translational fidelity by 
contributing an amino acid which is different from that dictated by 
the mRNA template. Erroneous substitution of one amino acid by another 
is called {\it mis-sense} error. Pre-translational {\it mis-charging} 
of tRNA is not the only possible cause of mis-sense error. Erroneous 
selection of a correctly charged near-cognate or non-cognate tRNA, i.e., 
a co-translational {\it mis-reading} of a codon, also contributes to 
mis-sense error \cite{parker89,cochella05,zaher09,johansson08,rodnina12}. 

Thus, at the beginning of each elongation cycle the macromolecular 
complex consisting of the ribosome and accessory molecules select one of 
the four possible pathways indicated by the tRNA selected: (i) correctly 
charged cognate tRNA, (ii) incorrectly charged cognate tRNA, (iii) 
correctly charged near-cognate tRNA, and (iv) correctly charged non-cognate 
tRNA. Along each of these pathways the sequence of intermediate steps are 
identical although the molecular identities of the complexes are different. 
In other words, the network of mechano-chemical states consist of four distinct 
cycles that share the same initial state. 

The time taken by a ribosome to incorporate a single amino acid in the growing 
protein is also the duration of the ribosome's dwell at the corresponding codon on the 
mRNA template. The distribution of the dwell times (DDT) characterizes the 
intrinsic stochastic nature of the process of elongation of the nascent protein 
by the ribosome.
Here we develop a stochastic kinetic model for the elongation phase of translation 
capturing, within a single mathematical framework, all the four cycles that 
branch out from the initial state. Our model also distinguishes between the 
concentrations of four distinct types of tRNA molecules, namely, correctly 
charged cognate tRNA, incorrectly charged cognate tRNA, correctly 
charged near-cognate tRNA and correctly charged non-cognate tRNA.
Solving the corresponding master equations  
(a set of coupled ordinary differential equations), for an appropriate initial condition 
that captures the beginning of translation of a codon, we obtain the DDT of the 
ribosome. Moreover, using the steady-state solutions of these master equations 
we derive the exact analytical expression for the average velocity of the ribosome 
which is also the average rate of amino acid incorporation (i.e., average rate of 
elongation of the nascent protein) catalyzed by the ribosomal machinery. 
This expression is a generalization of the MM equation and, as we 
demonstrate explicitly, it reduces to the standard form of MM equation in the 
appropriate special limits of our model. Few graphical plots of the average 
rate of elongation, corresponding to some typical values of the rate constants, 
are presented to provide an intuitive understanding of the relative contributions 
of the four competing cycles.

Three of the four pathways originating from the initial state lead to translational 
error if the cycle is completed by adding an amino acid to the growing protein. 
Therefore, as a byproduct of our calculation, we also get exact expressions 
for the translation error. The more stringent is the mechanism of selection of 
incoming tRNA the lower is the mis-reading error. But, increasing the probability 
of rejecting near-cognate and non-cognate tRNAs would increase the 
likelihood of accepting not only correctly charged cognate tRNA but also that 
of a mis-charged cognate tRNA. A few illustrative plots display the interplay of 
the effects of micharging of tRNA and misreading of mRNA in the overall 
mis-sense error in translation.

\section{Model}

\begin{figure}[h]
\begin{center}
\includegraphics[width=0.9\columnwidth]{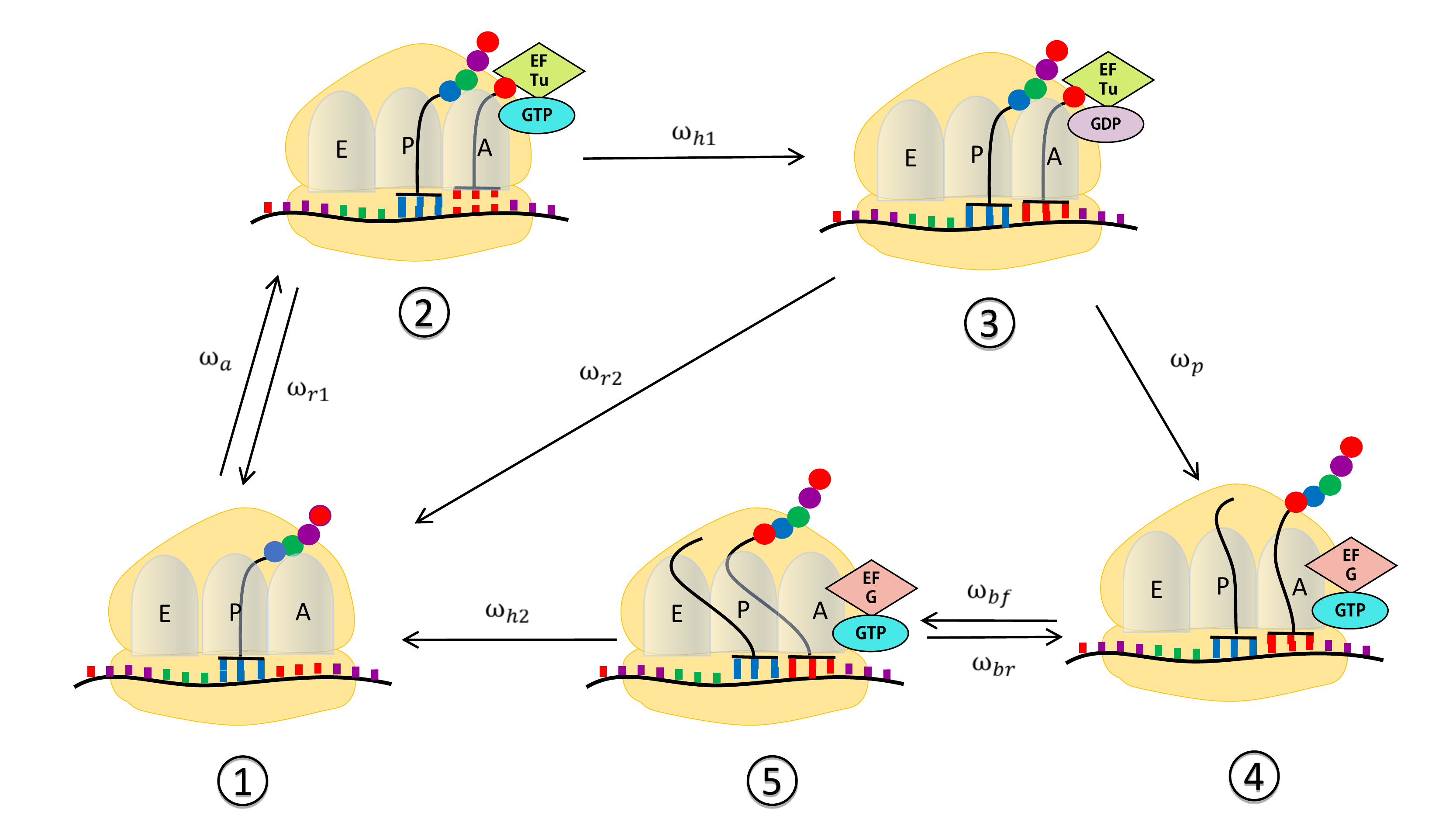}
\end{center}
\caption{Pictorial depiction of the elongation cycle in the SC model. 
(see the text for details).
\label{fig-SCmodel}}
\end{figure}

Sharma and Chowdhury \cite{sharma10} developed a 5-state stochastic 
kinetic model (from now onwards referred to as SC model) for the 
elongation cycle of translation (see Fig.\ref{fig-SCmodel}). 
The arrival of a aa-tRNA molecule, bound to GTP and EF-Tu, and its 
rejection because of the codon-anticodon mismatch are captured by the 
forward and reverse transitions $ 1 \rightleftharpoons 2$. 
The second stage of quality control (kinetic proofreading) involves 
hydrolysis of GTP by EF-Tu ($2 \to 3$) followed by either rejection 
($3 \to 1$) or incorporation ($3 \to 4$) in the growing protein by 
formation of a peptide bond (see ref.\cite{blomberg} for a pedagogical 
introduction to kinetic proofreading). The first of the two-step translocation 
process consists of the Brownian rotation of large subunit of the 
ribosome with respect to the small subunit and simultaneous reversible 
transitions of the tRNAs between the so-called ``classical'' and ``hybrid'' 
states. The second, and the final, step of translocation, driven by hydrolysis 
of another molecule of GTP by EF-G completes the cycle irreversibly. 
More detailed stochastic models of mechano-chemical kinetics of each 
elongation cycle have been developed (see, for example, 
\cite{xie13,thompson15}). However, in order to capture some other aspects of 
translational kinetics, we describe the kinetics of elongation cycle 
by the simpler SC model. Nevertheless, the strategy of modeling followed 
here can be implemented also using the more detailed descriptions as 
the basic models of elongation cycle.

The SC model was used further to account 
for the stochastic alternating pause-and-translocation kinetics of a 
single ribosome \cite{sharma11a} as well as for analyzing collective 
spatio-temporal organization of ribosomes in a polysome \cite{sharma11b}. 
Because of the extreme simplicity of the SC model model, no clear 
distinction could be made, in terms of different rate constants, between 
processes involving near-cognate and non-cognate tRNAs. More importantly, 
the SC model captured the possibility of mis-sense error arising from 
only mis-reading of the codons; it was not possible to incorporate the 
contributions from both mis-reading and mis-charging errors explicitly. 
The non-trivial extension of the SC model that we present here does not 
suffer from any of the above mentioned limitations of the original SC model.

\begin{figure}[h!]
\begin{center}
\includegraphics[width=0.9\columnwidth]{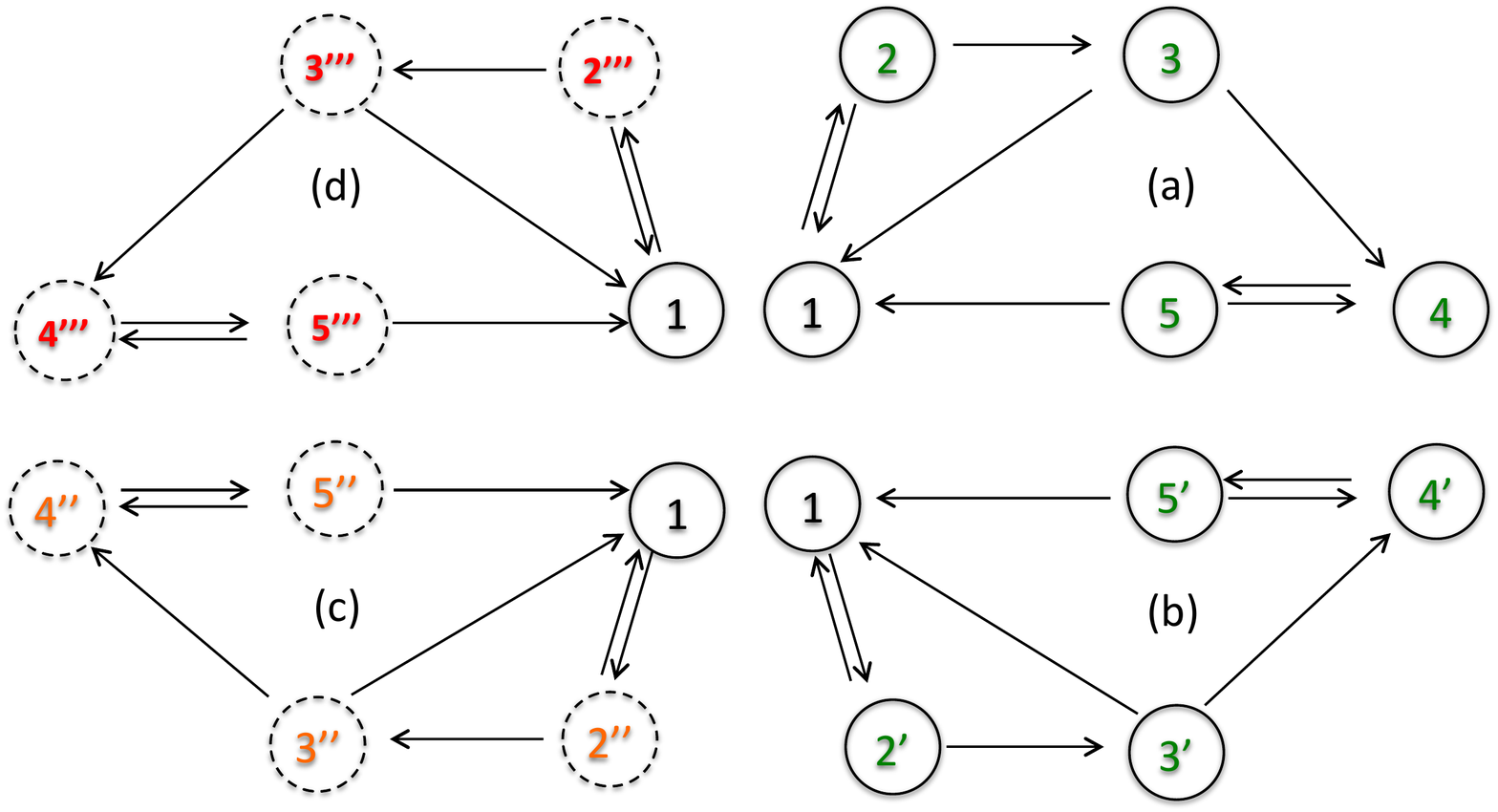}
\end{center}
\caption{Pictorial depiction of the four possible alternative mutually exclusive pathways that open up, in each chemo-mechanical cycle of a single ribosome, upon the arrival of (a) correctly charged cognate aa-tRNA, (b) incorrectly charged cognate aa-tRNA, (c) correctly charged near-cognate tRNA, and (d) correctly charged non-cognate tRNA (see the text for details). 
\label{fig-4cycles}}
\end{figure}

We begin formulation of the model with the four alternative elongation 
cycles shown in Fig.\ref{fig-4cycles} which correspond to the four 
different mutually exclusive pathways that open up with the arrival of 
(a) correctly charged cognate tRNA, (b) incorrectly charged cognate tRNA, 
(c) correctly charged near-cognate tRNA, and (d) correctly charged 
non-cognate tRNA. Note that each of these cycles is formally identical 
to the only cycle that appeared in the original SC model. However, by 
opening up the possibility of four distinct pathways, each associated 
with a distinct identity of aa-tRNA, this model not only allows for a 
clear distinction between non-cognate, near-cognate and cognate tRNAs 
but also that between correctly and incorrectly charged cognate tRNAs.

Next we simplify the model by exploiting some well known facts from 
the existing literature \cite{spirin02,rodnina11,frank11}. 
First, we note that $\omega_{a}, \omega_{a}', \omega_{a}''$ and $\omega_{a}'''$ 
are proportional to the concentrations of the corresponding aa-tRNA species; 
therefore, we assume: 
\begin{eqnarray}
\omega_{a} &=& \omega_{a\, c1}^{0}[\text{tRNA}]_{c1} \;\text{(correctly charged cognate tRNA)}\nonumber\\
\omega_{a}' &=& \omega_{a\, c2}^{0}[\text{tRNA}]_{c2} \;\text{(incorrectly charged cognate tRNA)}\nonumber\\
\omega_{a}'' &=& \omega_{a\, n}^{0}[\text{tRNA}]_{n} \;\text{(correctly charged near-cognate tRNA)}\nonumber\\
\omega_{a}''' &=& \omega_{a\, N}^{0}[\text{tRNA}]_{N} \;\text{(correctly charged non-cognate tRNA)}\nonumber\\
\label{eq-trnaconc}
\end{eqnarray} 
where the symbol [.] denotes the concentration of the corresponding 
tRNA species and the prefactors are measures of the intrinsic rates 
of the reactions for unit concentration of the tRNA species. Thus, as stated in the introduction, 
concentrations of all the four types of tRNA molecules are incorporated explicitly.

\begin{figure}[h]
\begin{center}
\includegraphics[width=0.9\columnwidth]{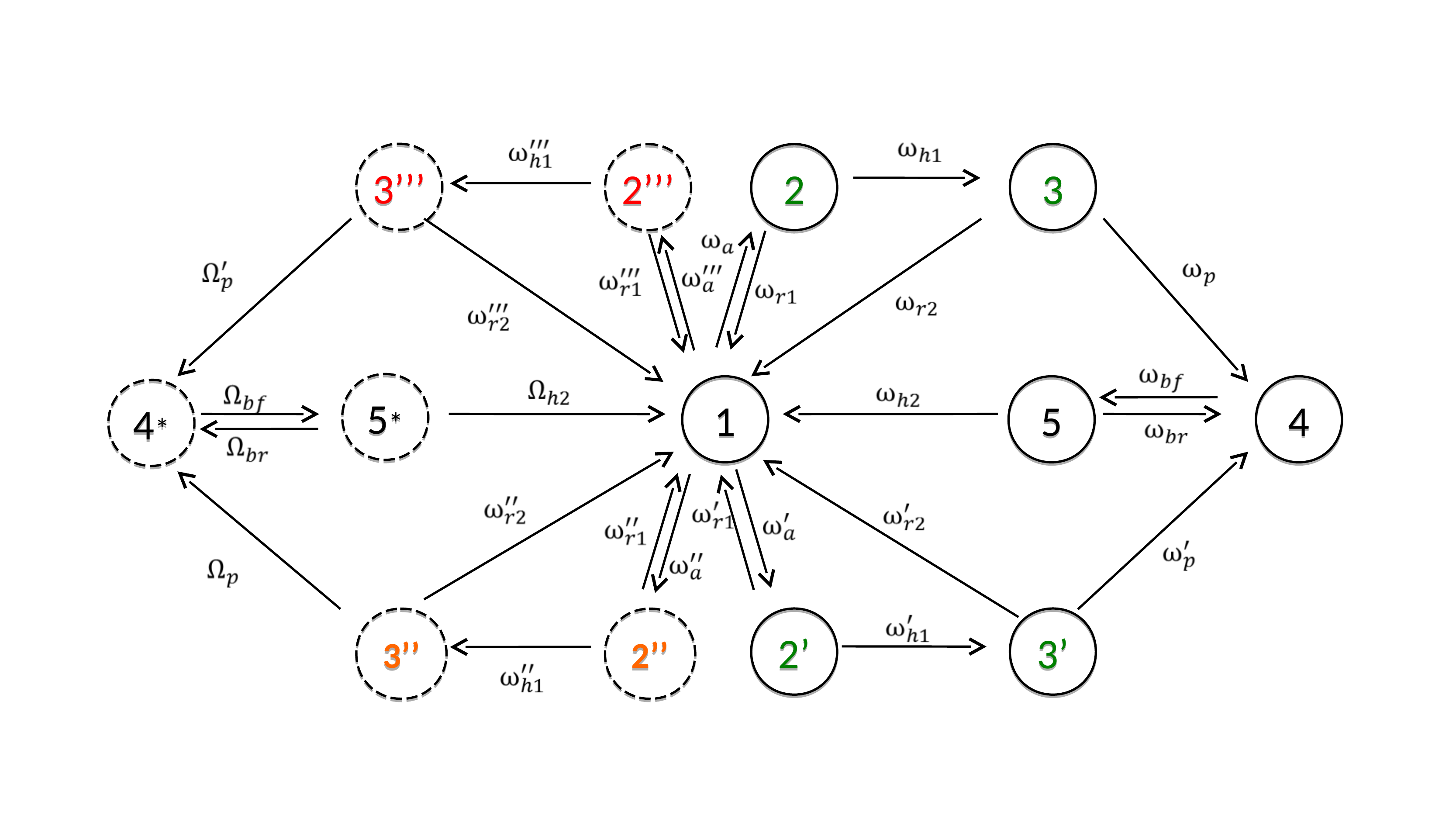}
\end{center}
\caption{Pictorial depiction of the full chemo-mechanical kinetics in the elongation cycle 
of a single ribosome, along with the corresponding rate constants. It is obtained from 
Fig.\ref{fig-4cycles} by combining the four cycles (see the text for details). 
\label{fig-ribomodel}}
\end{figure}

The assumption (\ref{eq-trnaconc}) is valid under the ``abundant substrate'' condition, i.e., 
all four species of tRNA molecules are much more abundant than the ribosomes. This condition 
is commonly used in the stochastic models of enzyme kinetics although strong deviation from 
this condition can lead to drastically different results \cite{grima17}.

We do not distinguish $4'$ from $4$ and $5'$ from $5$ because both the 
pathways $3 \to 4 \to 5$ and $3' \to 4' \to 5'$ involve movement of 
cognate tRNAs (see Fig.\ref{fig-ribomodel}). 
Similarly, assuming the rates of translocation of near-cognate and 
noncognate tRNA molecules to be comparable, but discriminating these 
from the corresponding cognate tRNAs, we assume 
$4'' \equiv 4'''=4^{*} \not \equiv 4$ and 
$5'' \equiv 5'''=5^{*} \not \equiv  5$ (see Fig.\ref{fig-ribomodel}). 
These assumptions help in combining the four pathways shown in 
Fig.\ref{fig-4cycles} within the single and simpler kinetic scheme 
depicted in Fig.\ref{fig-ribomodel} thereby also reducing the number 
of parameters (rate constants). From now onwards, unless stated otherwise, 
all our discussions will be based on the model kinetic scheme shown 
in Fig.\ref{fig-ribomodel}.

We use the symbol $P_{\mu}(j,t)$ to denote the probability at time $t$ that the ribosome is in the ``chemical'' state $\mu$ and is decoding the $j^{th}$ codon. In the steady state, all the probabilities 
$P_{\mu}(j,t)$ become independent of time. We define {\it translational fidelity} by the fraction  
\begin{eqnarray}
\phi &=& \dfrac{\omega_{p} P_{3}}{\omega_{p} P_{3} + \omega^{'}_{p} P^{'}_{3} + \Omega_{p} P^{''}_{3} + \Omega_{p}^{'} P^{'''}_{3}} = \dfrac{\omega_{p} P_{3}}{\omega_{p} P_{3} + \omega^{'}_{p} P^{'}_{3}      + \Omega_{h2} P^{*}_{5}} 
\end{eqnarray}
where we have used the relation 
$\Omega_{p} P^{''}_{3}+\Omega^{'}_{p} P^{'''}_{3}=\Omega_{h2} P^{*}_{5}$.

The {\it total mis-sense} error $E=1-\phi$ is defined by the relation
\begin{equation}
E =\dfrac{\omega'_{p} P'_{3}+ \Omega_{h2} P^{*}_{5}}{\omega_{p} P_{3} + \omega'_{p} P'_{3}+ \Omega_{h2} P^{*}_{5}} 
\end{equation}
which is the sum of the total {\it mis-charged mis-sense} error (i.e., mis-sense error arising solely from 
mis-charged tRNAs) 
\begin{equation}
E_{mc}=\dfrac{\omega'_{p} P'_{3}}{\omega_{p} P_{3} + \omega'_{p} P'_{3}+ \Omega_{h2} P^{*}_{5}} 
\end{equation}
and the total {\it mis-reading mis-sense} error (i.e., mis-sense error arising only from mis-reading 
of codons) 
\begin{equation}
E_{mr} =\dfrac{\Omega_{h2} P^{*}_{5}}{\omega_{p} P_{3} + \omega'_{p} P'_{3}+ \Omega_{h2} P^{*}_{5}}.  
\end{equation}
Similarly, the fraction 
\begin{eqnarray}
\epsilon_{mc} = \dfrac{\omega^{'}_{p} P^{'}_{3}}{\omega^{'}_{p} P^{'}_{3} + \Omega_{h2} P^{*}_{5}} 
\end{eqnarray} 
is the {\it fraction of mis-sense error caused by mis-charged cognate tRNAs}, 
while the corresponding fraction of mis-sense error caused by misreading is defined by
\begin{equation}
\epsilon_{mr}=\dfrac{\Omega_{h2} P^{*}_{5}}{\omega'_{p} P'_{3} + \Omega_{h2} P^{*}_{5}} 
\end{equation}

Obviously, the {\it average velocity} of a ribosome in the steady-state can be obtained by 
substituting the  expressions of ${\cal P}_{5}$ and ${\cal P}_5^*$ into the defining relation
\begin{equation}
V= {\ell}_c (\omega_{h2} {\cal P}_5 + \Omega_{h2} {\cal P}_5^*) 
\label{eq-v}
\end{equation} 
where ${\ell}_c$ is the length of a codon. We also note that the average velocity $V$ of a 
ribosome is same as the average rate of elongation of the protein that it polymerizes.  

We define the {\it rejection factors}
\begin{eqnarray}
{\cal R} &=& \biggl(\frac{\omega_{r1}}{\omega_{r1}+\omega_{h1}} \biggr)\biggl(\frac{\omega_{r2}}{\omega_{r2}+\omega_{p}} \biggr) \nonumber \\
{\cal R}' &=& \biggl(\frac{\omega'_{r1}}{\omega'_{r1}+\omega'_{h1}} \biggr)\biggl(\frac{\omega'_{r2}}{\omega'_{r2}+\omega'_{p}} \biggr) \nonumber \\
{\cal R}'' &=& \biggl(\frac{\omega''_{r1}}{\omega''_{r1}+\omega''_{h1}} \biggr)\biggl(\frac{\omega''_{r2}}{\omega''_{r2}+\Omega_{p}} \biggr) \nonumber \\
{\cal R}''' &=& \biggl(\frac{\omega'''_{r1}}{\omega'''_{r1}+\omega'''_{h1}} \biggr)\biggl(\frac{\omega'''_{r2}}{\omega'''_{r2}+\Omega'_{p}} \biggr) \nonumber \\
\label{eq-4rejectfactors}
\end{eqnarray}
The four rejection factors characterize the frequencies of rejection of the incoming charged tRNA molecules 
in the four alternative pathways depicted in Fig.\ref{fig-ribomodel}. The higher the value of a rejection factor 
the more frequent is the corresponding futile cycles.

The analytical results for this model that we report here are exact, i.e., these are derived 
without making any mathematical approximations. The derivations of these analytical 
expressions do not require imposition of any condition on the numerical values of the 
rate constants. However, we now list some biologically motivated constraints on the relative 
magnitudes of the rate constants that we'll use later in this paper only for presenting the 
results graphically for biologically relevant situations. 
Based on the levels of base-pair complementarity between the codon and the 
anticodon of the incoming tRNA, we expect that under normal physiological conditions the 
following conditions would be satisfied: 
$\omega_{r1}''' > \omega_{r1}'' > \omega_{r1}' = \omega_{r1}$. Motivated by similar 
considerations, for graphical plots, we also assume 
$\omega_{r2}''' > \omega_{r2}'' > \omega_{r2}' = \omega_{r2}$. 
Continuing similar justification for the reduction in the number of model parameters, we assume 
$\Omega_{p} \simeq \Omega_{p}' \simeq \omega_{p}'  < \omega_{p}$.

\section{Results} 

We begin our theoretical analysis by first solving the master equations (\ref{eq-master}) 
under steady-state conditions to get the corresponding expressions for $P_{\mu}$; the 
full analytical expressions are given in appendix \ref{sec-ssprobs}. Then using those 
expressions for $P_{\mu}$  we calculate the quantities of our 
interest namely, $\phi$, $E_{mc}$, $E_{mr}$, $\epsilon_{mc}$, $\epsilon_{mr}$ and $V$.
The results are listed below. 
\begin{eqnarray}
\phi= \dfrac{A}{A+B+C+D}
\label{eq-phiexp}
\end{eqnarray} 
and, hence,
\begin{eqnarray}
E = 1-\phi = \dfrac{B+C+D}{A+B+C+D} 
\label{eq-Eexp}
\end{eqnarray}
which is sum of the the two contributions
\begin{eqnarray}
E_{mc} = \dfrac{B}{A+B+C+D}
\label{eq-Emcexp}
\end{eqnarray}
and
\begin{equation}
E_{mr}=\dfrac{C+D}{A+B+C+D}.
\label{eq-Emrexp}
\end{equation}
Similarly, we get 
\begin{eqnarray}
\epsilon_{mc} = \dfrac{B}{B+C+D}
\label{eq-epmcexp}
\end{eqnarray} 
and
\begin{eqnarray}
\epsilon_{mr} = 1-\epsilon_{mc} = \dfrac{C+D}{B+C+D}
\label{eq-epmrexp}.
\end{eqnarray}
In all the expressions  (\ref{eq-phiexp})-(\ref{eq-epmrexp}) $A, B, C$ and $D$ are given by 
\begin{eqnarray}
A &=& \frac{\omega _a}{\left[1+(\omega _{\text{r1}}/\omega _{\text{h1}})\right] \left[1+(\omega _{\text{r2}}/\omega _p)\right]} 
\nonumber \\
B &=& \frac{\omega _a'}{\left[1+(\omega _{\text{r1}}'/\omega _{\text{h1}}')\right] \left[1+(\omega _{\text{r2}}'/\omega _p')\right]}
\nonumber \\
C &=& \frac{\omega _a''}{\left[1+(\omega _{\text{r1}}''/\omega _{\text{h1}}'')\right] \left[1+(\omega _{\text{r2}}''/\Omega _p)\right]}
\nonumber \\
D &=& \frac{\omega _a'''}{\left[1+(\omega _{\text{r1}}'''/\omega _{\text{h1}}''')\right] \left[1+(\omega _{\text{r2}}'''/\Omega _p')\right]}
\end{eqnarray}

The expressions (\ref{eq-phiexp})-(\ref{eq-epmrexp}) can be easily justified by intuitive 
physical arguments. Let us first consider the special case where 
$\omega_{\text{r1}} = \omega_{\text{r1}}' = \omega_{\text{r1}}'' = \omega_{\text{r1}}''' = 0 
=\omega_{\text{r2}} = \omega_{\text{r2}}' = \omega_{\text{r2}}'' = \omega_{\text{r2}}'''$. 
In this case the expressions for $A, B, C$ and $D$ reduce to 
$A=\omega_{a},  B=\omega_{a}', C=\omega_{a}''$ and $D=\omega_{a}'''$, respectively. 
Consequently, $\phi = \omega_{a}/(\omega_{a}+\omega_{a}'+\omega_{a}''+\omega_{a}''')$ 
is the probability of following the path $1 \to 2$, instead of the other three alternatives, 
namely, $1 \to 2'$, $1 \to 2''$ and $1 \to 2'''$. Similarly, in this special case, the expression
$\epsilon_{mc} = \omega_{a}'/(\omega_{a}'+\omega_{a}''+\omega_{a}''')$ is expected 
because $\epsilon_{mc}$ is the probability of following the path $1 \to 2'$, instead of the 
two alternatives $1 \to 2''$ and $1 \to 2'''$.

In the general case, the {\it rate} of transition $1 \to 2 \to 3 \to 4$ is given by 
\begin{equation}
\omega_{a}~~ \underbrace{\frac{\omega_{h1}}{\omega_{h1}+\omega_{r1}}}_{\text{Prob. for $2 \to 3$ }}~~  \underbrace{\frac{\omega_{p}}{\omega_{p}+\omega_{r2}}}_{\text{Prob. for $3 \to 4$ }} = A
\end{equation}
The quantities $B$, $C$ and $D$ have similar interpretations as rates for 
the transitions $1 \to 2' \to 3' \to 4$, $1 \to 2'' \to 3'' \to 4^{*}$ and $1 \to 2''' \to 3''' \to 4^{*}$, 
respectively. Once the system reaches the state $4$ it cannot return to state $1$ without completing 
the full cycle. Therefore, fidelity $\phi$ is the ratio $A/(A+B+C+D)$. The expressions 
(\ref{eq-Emcexp})-(\ref{eq-epmrexp}) for $E_{mc}, E_{mr}$ and $\epsilon_{mc}, \epsilon_{mr}$ 
also follow from the same interpretations of $A, B, C$ and $D$.

\begin{figure}[h!]
\begin{center}
\includegraphics[width=0.9\columnwidth]{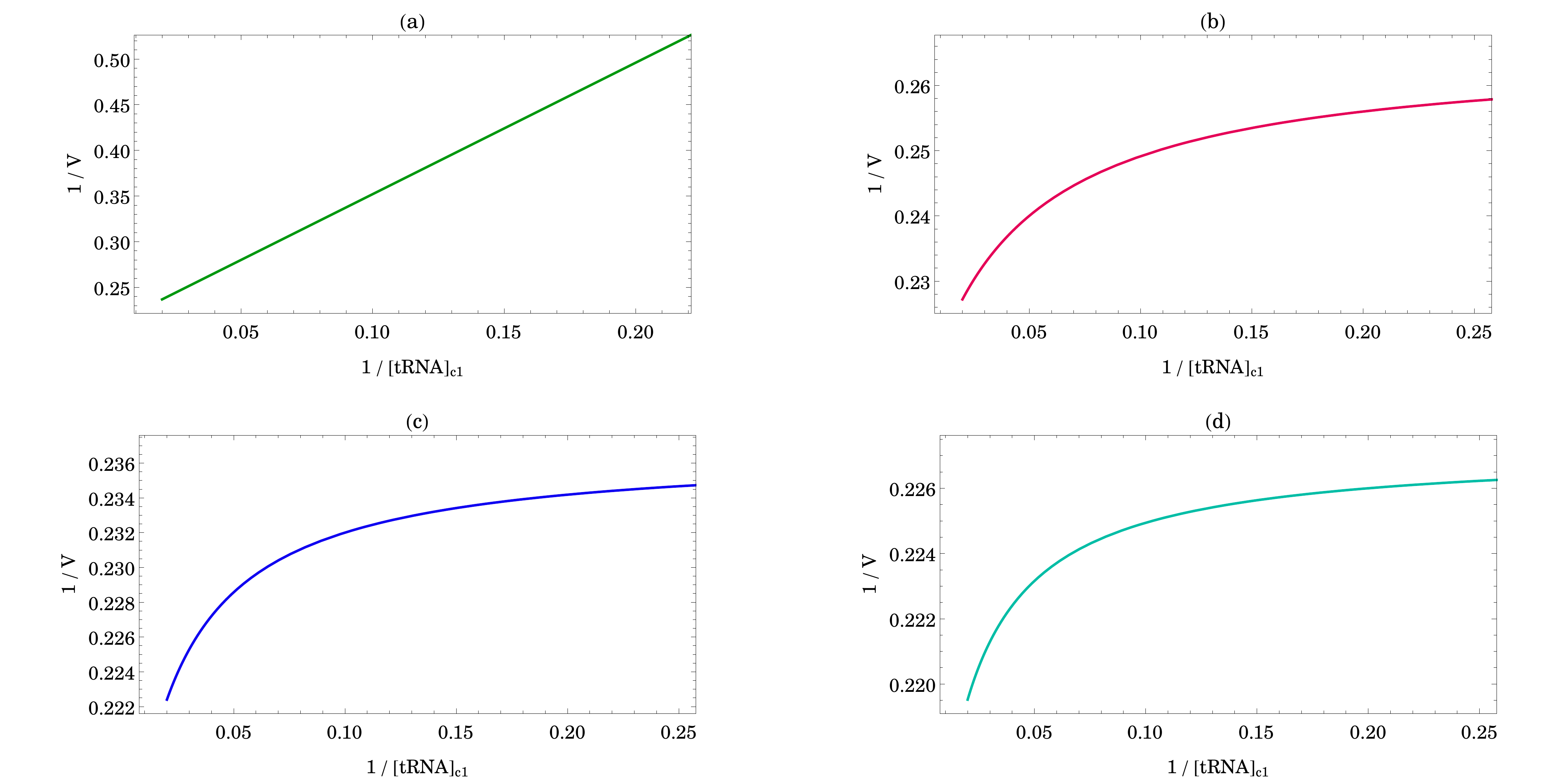}
\end{center}
\caption{$1/V$ is plotted against $1/[tRNA]_{c1}$ for our model; this plot is the analog of  Lineweaver-Burk plot for the Michaelis-Menten reaction. For all the four plots (a)-(d), except for $\omega_{a}$, $\omega'_{a}$, $\omega''_{a}$, $\omega'''_{a}$, the numerical values assigned to the rate constants for all the four tRNA species are those listed in the {\it first column} of TABLE I while $\omega_{a}$ is varied from  $0 \; to \; 50 \; s^{-1}$. The other parameters are as follows: (a) $\omega'_{a}\;=\;0$, $\omega''_{a}\;=\;0$ and $ \omega'''_{a}\;=\;0$; (b) $\omega'_{a}\;=\;25\; s^{-1}$, $\omega''_{a}\;=\;0$ and $\omega'''_{a}\;=\;0$; (c) $\omega'_{a}\;=\;25\; s^{-1}$, $\omega''_{a}\;=\;25\; s^{-1}$ and $ \omega'''_{a}\;=\;0$; and (d) $\omega'_{a}\;=\;25\; s^{-1}$, $\omega''_{a}\;=\;25\; s^{-1}$ and $\omega'''_{a}\;=\;25\; s^{-1}$.
}
\label{fig-MMlike}
\end{figure}

Ribosome is an enzyme; interestingly, at any given instant of time its substrate-specificity 
depends on the codon that it is engaged in translating. In recent years, the average rate 
of translation has been shown to be a generalization of the rate of enzymatic reactions.
Recall that for the Michaelis-Menten (MM) enzymatic reaction 
\begin{equation}
E + S \mathop{\rightleftharpoons}^{k_{+1}}_{k_{-1}}  [ES] \mathop{\rightarrow}^{k_{2}} E + P 
\label{eq-MMreac} 
\end{equation} 
the rate of the reaction under steady-state condition is given by the MM equation 
\begin{equation}
\frac{1}{V} = \frac{1}{V_{max}} + \frac{K_M}{V_{max}}\frac{1}{[S]}
\label{eq-MMeq}
\end{equation}
where the Michaelis constant $K_{M} = (k_{-1}+k_{2})/k_{+1}$ and  $V_{max}=k_{2} [E]_{0}$, 
with $[E]_{0}$ being the initial (total) concentration of the enzyme. In the past the average rate of 
translation by a ribosome have been shown to follow a generalized MM-like equation where 
the concentration of aa-tRNA is interpreted as the substrate concentration. 
For simpler models of translation reported earlier, the average rate of translation has been 
expressed as generalized MM equation \cite{garai09,chowdhury14}.

For the full kinetic model shown in Fig.\ref{fig-ribomodel} the average rate of translation 
(i.e., the average velocity $V$  of a ribosome) is given by 
\begin{eqnarray}
\dfrac{1}{V}&=&\dfrac{1}{A+B+C+D} \nonumber \\ 
&+& \dfrac{A}{A+B+C+D} \biggl(\frac{1}{V_{A}}\biggr) 
+ \dfrac{B}{A+B+C+D} \biggl(\frac{1}{V_{B}}\biggr) + \dfrac{C}{A+B+C+D} \biggl(\frac{1}{V_{C}}\biggr)  
+ \dfrac{D}{A+B+C+D} \biggl(\frac{1}{V_{D}}\biggr) 
\label{eq-onebyV}
\end{eqnarray}
where
\begin{eqnarray}
\frac{1}{V_{A}} &=&  \bigg [\dfrac{1}{\omega_{h1}}\bigg(1+\dfrac{\omega_{r2}}{\omega_p}\bigg)+\dfrac{1}{\omega_p}+\dfrac{1}{\omega_{bf}}\bigg(1+\dfrac{\omega_{br}}{\omega_{h2}}\bigg)+\dfrac{1}{\omega_{h2}} \bigg ] \nonumber \\
\frac{1}{V_{B}} &=& \bigg [\dfrac{1}{\omega_{h1}'}\bigg(1+\dfrac{\omega_{r2}'}{\omega_p'}\bigg)+\dfrac{1}{\omega_p'}+\dfrac{1}{\omega_{bf}}\bigg(1+\dfrac{\omega_{br}}{\omega_{h2}}\bigg)+\dfrac{1}{\omega_{h2}} \bigg ] \nonumber \\
\frac{1}{V_{C}} &=& \bigg [\dfrac{1}{\omega_{h1}''}\bigg(1+\dfrac{\omega_{r2}''}{\Omega_p}\bigg)+\dfrac{1}{\Omega_p}+\dfrac{1}{\Omega_{bf}}\bigg(1+\dfrac{\Omega_{br}}{\Omega_{h2}}\bigg)+\dfrac{1}{\Omega_{h2}} \bigg ] \nonumber \\
\frac{1}{V_{D}} &=&  \bigg [\dfrac{1}{\omega_{h1}'''}\bigg(1+\dfrac{\omega_{r2}'''}{\Omega_p'}\bigg)+\dfrac{1}{\Omega_p'}+\dfrac{1}{\Omega_{bf}}\bigg(1+\dfrac{\Omega_{br}}{\Omega_{h2}}\bigg)+\dfrac{1}{\Omega_{h2}} \bigg ]
\label{eq-onebyVa}
\end{eqnarray}
Eqn.(\ref{eq-onebyV}) is a generalized version of the MM equation (\ref{eq-MMeq}) for our model.
An intuitive derivation of the expression (\ref{eq-onebyV}), that provides a deeper physical 
interpretation of this formula, is given in appendix \ref{sec-cleland}.

The connection between the two can be elucidated by considering a special case of our model.
In the limit $[tRNA]_{c2} = [tRNA]_{n} = [tRNA]_{N} = \omega_{r2} = \omega_{br} = 0$ and 
$\omega_{p} \to \infty$, $\omega_{bf} \to \infty$, $\omega_{h2} \to \infty$, the model reduces 
to 
\begin{equation}
{\rm Ribosome}+tRNA_{c1} {\mathop{\rightleftharpoons}^{\omega^{0}_{a,c1}}_{\omega_{r1}}} 2 \mathop{\rightarrow}^{\omega_{h1}} 5. 
\end{equation} 
which is, formally, identical to the MM reaction (\ref{eq-MMreac}). In this limit the expression 
(\ref{eq-onebyV}) reduces to 
\begin{equation}
\frac{1}{V} = \frac{1}{\omega_{h1}} + \biggl(\frac{(\omega_{h1}+\omega_{r1})/\omega^{0}_{a,c1}}{\omega_{h1}}\biggr)\frac{1}{[tRNA]_{c1}} 
 \end{equation} 
which is identical to MM equation (\ref{eq-MMeq}) because of the correspondence 
$\omega_{h1} \longleftrightarrow k_{2} = V_{max}$ and 
$(\omega_{h1}+\omega_{r1})/\omega^{0}_{a,c1} \longleftrightarrow (k_{-1}+k_{2})/k_{1} = K_{M}$.
Thus, on the right hand side of the equation (\ref{eq-onebyV}) the sum of the last four terms is 
the generalized counterpart of $1/V_{max}$ while the first term is the generalized analog of 
$K_{M}/V_{max}[S]$.

\begin{figure}[h]
\begin{center}
\includegraphics[width=0.9\columnwidth]{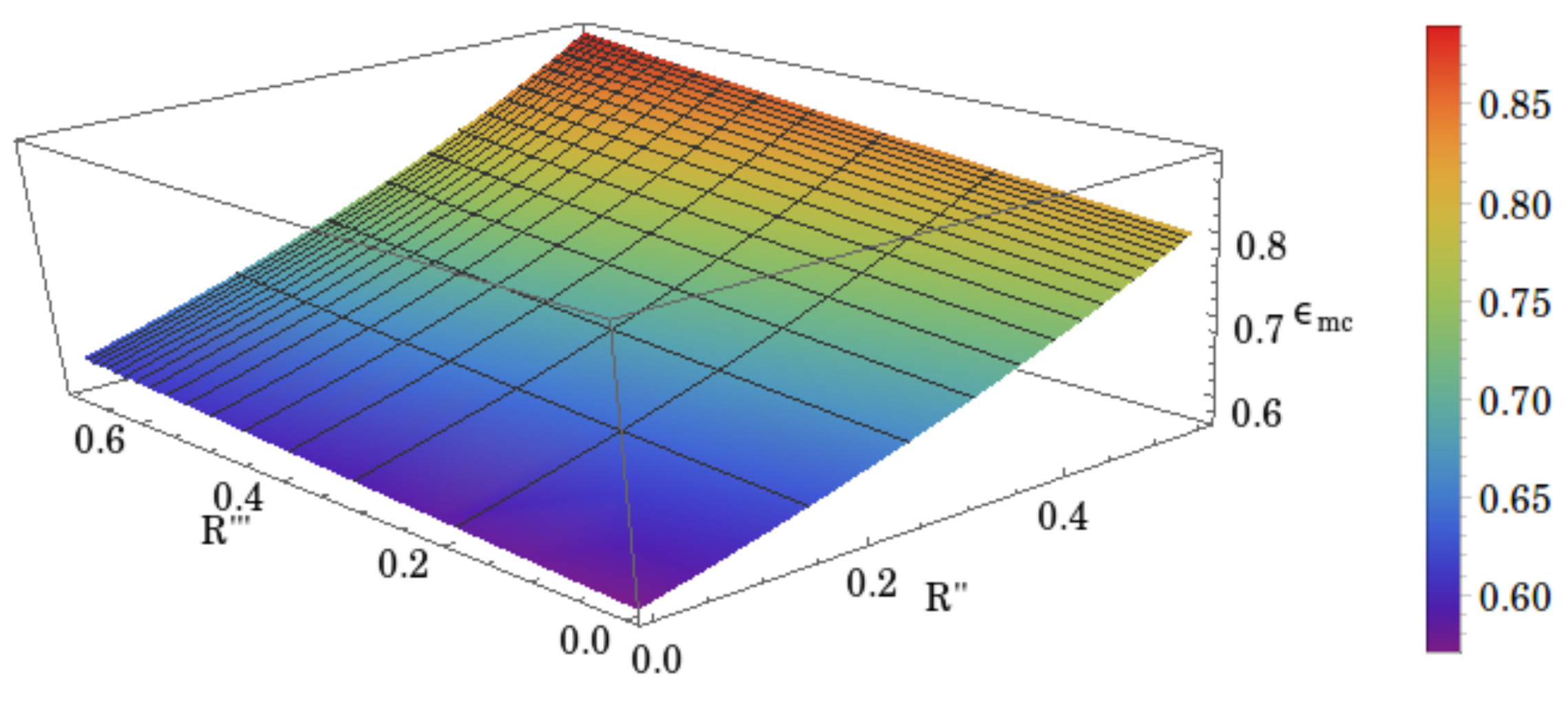}
\end{center}
\caption{The fraction $\epsilon_{mc}$ of error caused by mis-charged 
tRNA is plotted in 3D against the rejection factors ${\cal R}''$ and 
${\cal R}'''$ of the near-cognate and non-cognate tRNAs, respectively. 
The parameters $\omega_{r2}''$ and $\omega_{r2}'''$ have been varied 
from $0$ to 50 s$^{-1}$ keeping the values of all the other parameters 
fixed at those listed in table \ref{tab-table1}.
}
\label{fig-Figdata1}
\end{figure}

\begin{figure}[h]
\begin{center}
\includegraphics[width=0.9\columnwidth]{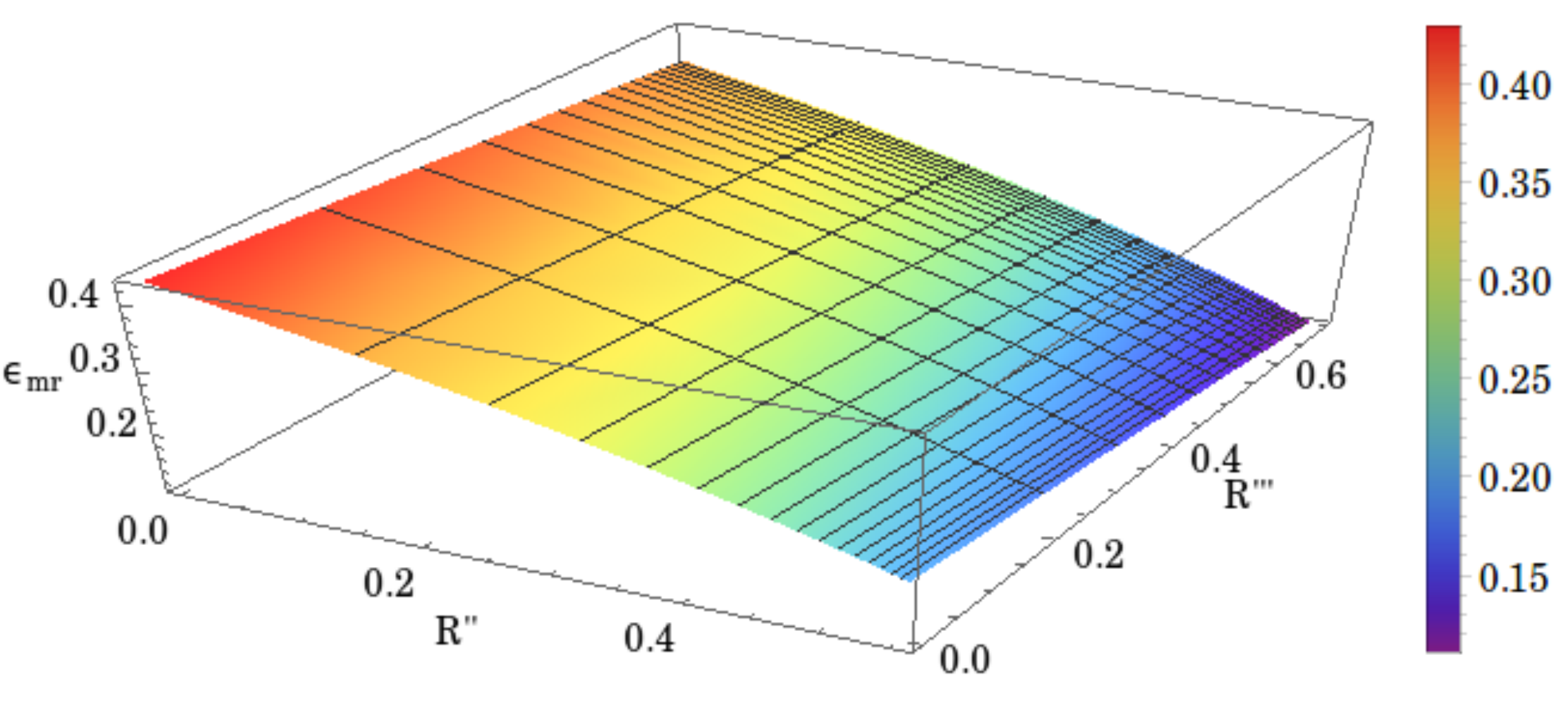}
\end{center}
\caption{Same as in Fig.\ref{fig-Figdata1}, except that the fraction 
$\epsilon_{mr}$ of error caused by misreading of mRNA is plotted 
against ${\cal R}''$ and ${\cal R}'''$.  
}
\label{fig-Figdata2}
\end{figure}

\begin{figure}[h]
\begin{center}
\includegraphics[width=0.8\columnwidth]{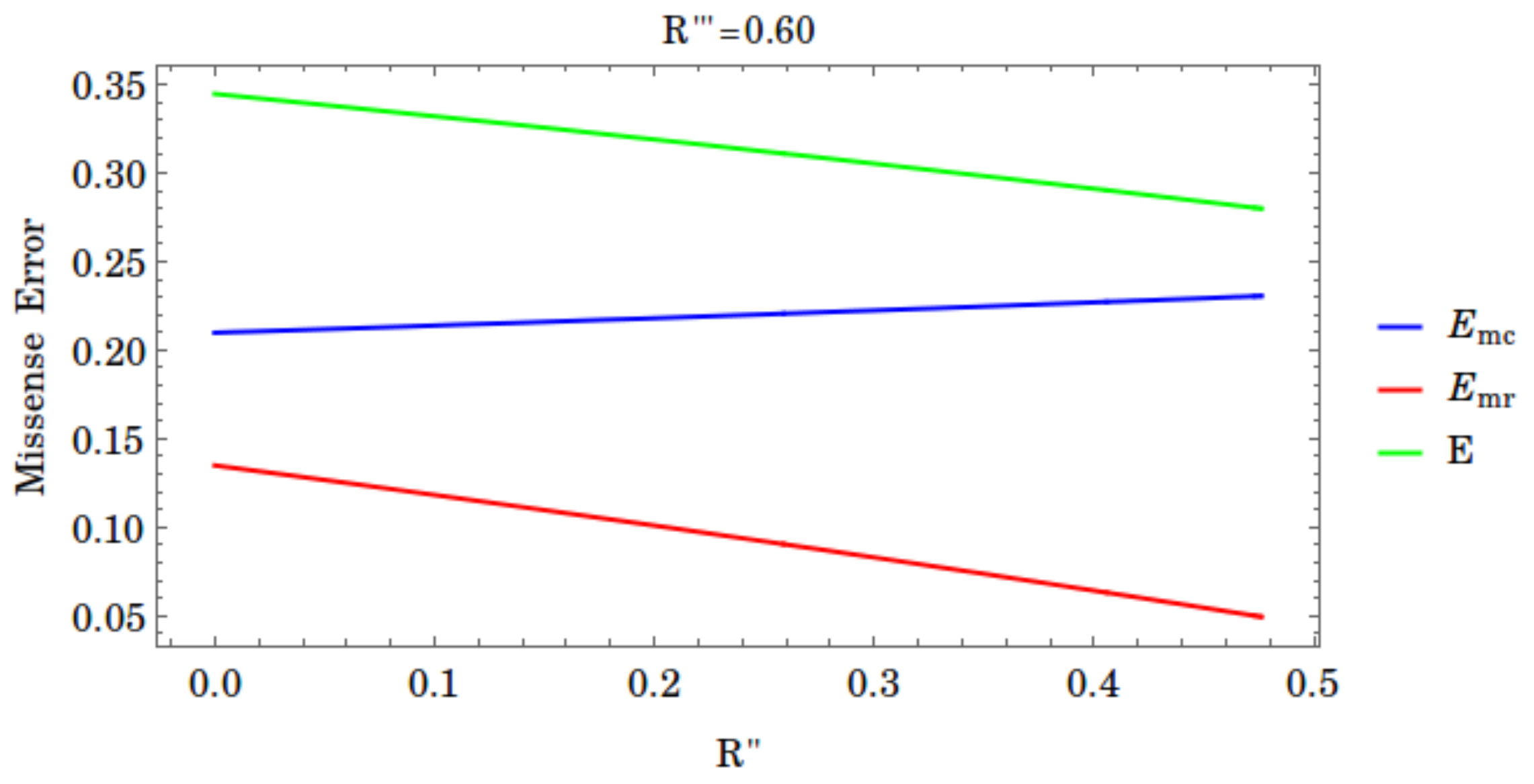}
\end{center}
\caption{The errors $E_{mc}$, $E_{mr}$ and $E$ are plotted against 
the rejection factor ${\cal R}''$ for a fixed value of ${\cal R}'''$ 
The parameter $\omega_{r2}''$ has been varied 
from $0$ to 25 s$^{-1}$ keeping the values of all the other parameters 
fixed at those listed in table \ref{tab-table1}.
}
\label{fig-Figdata3}
\end{figure}

Note that, from the perspective of enzymatic reactions, in each elongation cycle four 
distinct species of substrates (tRNA molecules) compete for the same enzyme. In order to 
graphically demonstrate the effects of this competition among the substrates, in 
Fig.\ref{fig-MMlike} we plot $1/V$ as a function of $1/[tRNA]_{c1}$ under four different conditions. 

The Fig.\ref{fig-MMlike}(a) corresponds to the simplest scenario where only correctly charged 
tRNA molecules are present. In this case, because of the absence competition among substrates, 
the plot is linear. This linear plot in fig.\ref{fig-MMlike}(a) is the characteristic of MM equation 
displayed in what is known as the  Lineweaver-Burk plot. 
By fitting the data of Fig. 4(a) to the MM equation (\ref{eq-MMeq}), where the 
substrate concentration $[S]$ is identified as  $[tRNA]_{c1}$,  we find $K_{m}/V_{max}=1.44$ and 
$1/V_{max}=0.208$; hence $K_{m} \simeq 6.9$ and $V_{max} \simeq 4.8$ amino acids per second 
(or, equivalently, codons per second). 
The deviation from the 
linearity, as shown in Fig.\ref{fig-MMlike}(b), arises from the presence of the competing second 
species of the tRNA molecules. The corresponding plots in Figs.\ref{fig-MMlike}(c) and (d) exhibit 
the increasing deviations from the single-substrate MM equation as the number of competing 
substrates increases. Interestingly, all the curves plotted in Figs.\ref{fig-MMlike} (a)-(d) have the 
same slope $1.44$ in the limit $1/[tRNA]_{c1} \to 0$, i.e., $[tRNA]_{c1} \to \infty$. One way of 
characterizing the extent of the deviation of the curves in Figs.\ref{fig-MMlike}(b)-(d) from linearity, 
with the increasing number of competing substrates, is by computing the saturation values of these 
curve in the limit $1/[tRNA]_{c1} \to \infty$; these values are approximately $ 0.27, 0.24$ and $0.23$, 
respectively.

The numerical values assigned to the rate constants for the plots in Fig.\ref{fig-MMlike} are 
not realistic in the sense that these do not reflect the intuitively expected relative strengths 
of the corresponding rates for the four different species of tRNA molecules. The sole purpose 
of treating all the four species of tRNA molecules on equal footing is to demonstrate the 
trend of increasing deviation from linearity on the  Lineweaver-Burk plot with the increasing 
number of competing substrates. 
For plotting all the remaining graphs, we have used the parameter  values as given in the table \ref{tab-table1}; .

\begin{table}[h]
\caption{Values for different parameters}
\begin{center}
\begin{tabular}{|c|c|c|c|}
\hline 
\multicolumn{1}{|p{2cm}}{\centering for correctly charged cognate tRNA}
& \multicolumn{1}{|p{2cm}}{\centering for incorrectly charged cognate tRNA}
& \multicolumn{1}{|p{2cm}}{\centering for near-cognate tRNA}
& \multicolumn{1}{|p{2cm}|}{\centering for non-cognate tRNA}\\
\hline
$\omega_{a}=$ 25 $s^{-1}$ & $\omega_{a}'=$ 10 $s^{-1}$ &$\omega_{a}''=$ 10 $s^{-1}$ &$\omega_{a}'''= $ 5 $s^{-1}$ \\
\hline
$\omega_{r1}=$ 5 $s^{-1}$ & $\omega_{r1}'=$ 5 $s^{-1}$ &$\omega_{r1}''=$ 20 $s^{-1}$ &$\omega_{r1}'''= $ 25 $s^{-1}$ \\
\hline 
$\omega_{r2}=$ 5 $s^{-1}$ & $\omega_{r2}'=$ 5 $s^{-1}$ &$\omega_{r2}''=$ 20 $s^{-1}$ &$\omega_{r2}'''= $ 25 $s^{-1}$ \\
\hline
$\omega_{h1}=$ 25 $s^{-1}$ & $\omega_{h1}'=$ 25 $s^{-1}$ &$\omega_{h1}''=$ 10 $s^{-1}$ &$\omega_{h1}'''= $ 5 $s^{-1}$ \\
\hline 
$\omega_{p}=$ 25 $s^{-1}$ & $\omega_{p}'=$ 10 $s^{-1}$ &$\Omega_{p}=$ 10 $s^{-1}$ &$\Omega_{p}'= $ 10 $s^{-1}$ \\
\hline
$\omega_{h2}=$ 25 $s^{-1}$ & $\omega_{h2}=$ 25 $s^{-1}$ &$\Omega_{h2}=$ 10 $s^{-1}$ &$\Omega_{h2}= $ 10 $s^{-1}$ \\
\hline 
$\omega_{bf}=$ 25 $s^{-1}$ & $\omega_{bf}=$ 25 $s^{-1}$ &$\Omega_{bf}=$ 10 $s^{-1}$ &$\Omega_{bf}= $ 10 $s^{-1}$ \\
\hline
$\omega_{br}=$ 25 $s^{-1}$ & $\omega_{br}=$ 25 $s^{-1}$ &$\Omega_{br}=$ 10 $s^{-1}$ &$\Omega_{br}= $ 10 $s^{-1}$ \\
\hline
\end{tabular}
\end{center}
\label{tab-table1}
\end{table}

The two-dimensional plots of the error fractions $\epsilon_{mc}$ and 
$\epsilon_{mr}$ against the rejection factors ${\cal R}''$ and 
${\cal R}'''$ are shown in Figs.\ref{fig-Figdata1} and \ref{fig-Figdata2}, 
respectively. Both show how the error fraction $\epsilon_{mr}$ 
decreases, while the fraction $\epsilon_{mc}$ increases with 
increasing ${\cal R}''$ and ${\cal R}'''$. The total mis-sense error 
can also decrease because, under favorable conditions, the increase 
of $E_{mc}$ with ${\cal R}''$ is more than compensated by the 
simultaneous decrease of $E_{mr}$, as shown in Fig.\ref{fig-Figdata3}. 

One of the main results reported above is the analytical expression for the average rate of translation, 
as given by Eq.\ref{eq-onebyV}, in the steady state. This average rate is the inverse of the mean time 
of dwell of a ribosome at successive codons \cite{sharma11a} in the stochastic model reported in this 
paper. Ideally, for any such stochastic model of translational kinetics it is desirable to derive the full 
probability density for the dwell times of a ribosome. Therefore, we now give an outline of our 
derivation of the probability density $f(t)$ of the dwell times of a ribosome in our model.

In order to simplify our calculations, following the trick used in ref.\cite{sharma11a}, we assume that the 
ribosome makes a transition to a hypothetical state $\tilde{1}$ at the $(j+1)$-th codon, after reaching 
the chemical state $5$ or $5^{*}$ at the $j$-th codon. It then relaxes to the chemical state 1 at the 
same codon $j+1$ at a rate $\delta$. We can recover our original model by taking $\delta\rightarrow \infty$. 
The probability of finding the ribosome in this hypothetical state is denoted by $\widetilde{P_1}(j+1,t)$. 
Now, we define the dwell time of the ribosome at a particular codon, say the $j$-th, by the time taken by 
the ribosome to reach  state $\tilde{1}$ at the $(j+1)$-th codon, starting from the state 1 at the the $j$-th 
codon. The master equations governing the time evolution of the probabilities, the normalization 
condition as well as the initial conditions for the set of master equations are given in Appendix C.
The probability of incorporation of one amino acid to the growing polypeptide in the time 
interval between t and $t+\Delta t$ is $f(t)\Delta t$ where $f(t)$ is given by \cite{sharma11a}
\begin{equation}
f(t)=\dfrac{\Delta\widetilde{P_{1}}(t)}{\Delta t}= \omega_{h2}P_{5}(t) + \Omega_{h2}P^{*}_{5}(t).
\label{eq-dwellTdefn}
\end{equation}

\begin{figure}[h]
\begin{center}
\includegraphics[scale=0.8]{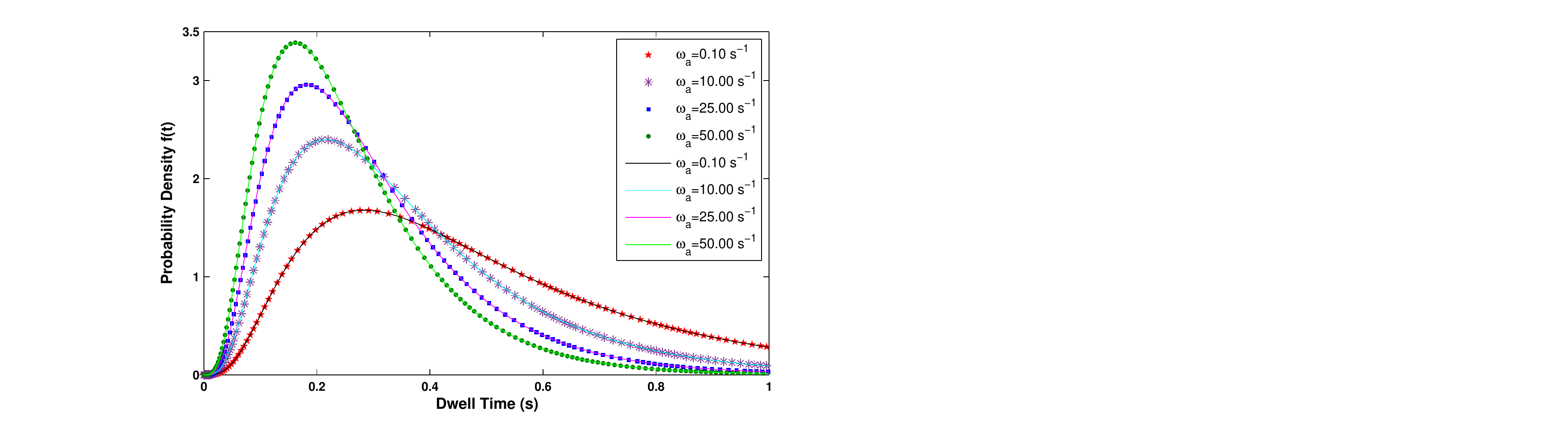}

\caption{{\bf The probability density of dwell times is plotted for four different values of $\omega_a$  keeping all the other parameters fixed at the values listed in table \ref{tab-table1}. The lines have been obtained by inverse Laplace transform of analytically derived expressions in Laplace space for the specific parameter values listed in table \ref{tab-table1} (see the text for details). The data obtained from the alternative direct numerical solution of the master equations (see the text for details) have been plotted using discrete symbols. } 
}
\label{fig-Figdatan}
\end{center}
\end{figure}

Since $P_{\mu}(t)$ are time-dependent solutions under the specific initial condition mentioned above, we cannot use the steady-state (i.e., time-independent) solutions derived in Appendix A. Instead, we adopted two alternative approaches for finding the probability density $f(t)$. In the first, which is essentially an analytical approach, we found $f(t)$ by the use of a standard technique \cite{sharma11a} based on Laplace transform. However, the analytical expressions of $P_{5}(s)$ and $P_{5}^{*}(s)$ in the Laplace space are too long (covering several pages) to be reproduced here. Therefore, instead, we have substituted the numerical values of the parameters listed in table \ref{tab-table1} and, then, carried out the inverse Laplace transform that involved finding the roots of a 5th degree polynomial which were calculated numerically. The four curves plotted graphically (by lines) in Fig.\ref{fig-Figdatan} have been obtained by repeating this procedure for four different values of $\omega_{a}$. In the second approach, for the given initial condition, we numerically solved the master equations given in appendix C (which are essentially a set of coupled ordinary differential equations) by a standard ODE solver in Matlab and hence obtained $f(t)$ by substituting 
$P_{5}(t)$ and $P^{*}_{5}(t)$ into (\ref{eq-dwellTdefn}). These numerical results for $f(t)$ are plotted in Fig.\ref{fig-Figdatan} by discrete symbols. Results obtained by the two methods are in excellent agreement with each other.

As $\omega_{a}$ increases the probability density $f(t)$ becomes sharper. This trend of variation of the width of the distribution is consistent with the intuitive expectation that fluctuations in the dwell time, caused by the low concentration of tRNA, would become stronger with the decrease of $\omega_{a}$ which is directly proportional to $[tRNA]_{c1}$. To quantify the relative strength of the fluctuation and mean of the dwell times we have computed the numerical values of the randomness parameter, defined as 
\begin{equation}
r=\frac{<t^{2}>-<t>^{2}}{<t>^{2}}, 
\end{equation}
for the four curves plotted in  Fig.\ref{fig-Figdatan}; these data are presented in table II.

\begin{table}[h]
\def\arraystretch{1.5}
\begin{center}
\begin{tabular}{|c|c|} 
\hline
$\omega_a$ (s$^{-1}$) & Randomness Parameter (r) \\
\hline
0.1 & 0.434293 \\
\hline
10.0 & 0.392075  \\
\hline 
25.0 & 0.371798 \\
\hline 
50.0 & 0.363056 \\
\hline 
\end{tabular}
\caption{\bf The numerical values of the randomness parameter $r=(<t^{2}>-<t>^{2})/<t>^{2}$ for the four curves plotted in Fig.\ref{fig-Figdatan}}
\end{center}
\label{tab-table4}
\end{table}

Naturally occuring mRNA templates in living cells consist of a heterogeneous 
sequence of nucleotides and, consequently, not all the codons are identical, 
in general. This feature of real mRNA templates can be captured in our 
theoretical model by assigning to each of rate constant different numerical 
values for translating different codons. However, only numerical results can 
be obtained in such cases. But, in order to derive the results  analytically 
(in terms of closed form mathematical formulae), in this paper we have 
considered the special case where the numerical values of each rate constant 
is independent of the type of the codon.

\begin{figure}[h]
\begin{center}
\includegraphics[width=0.9\columnwidth]{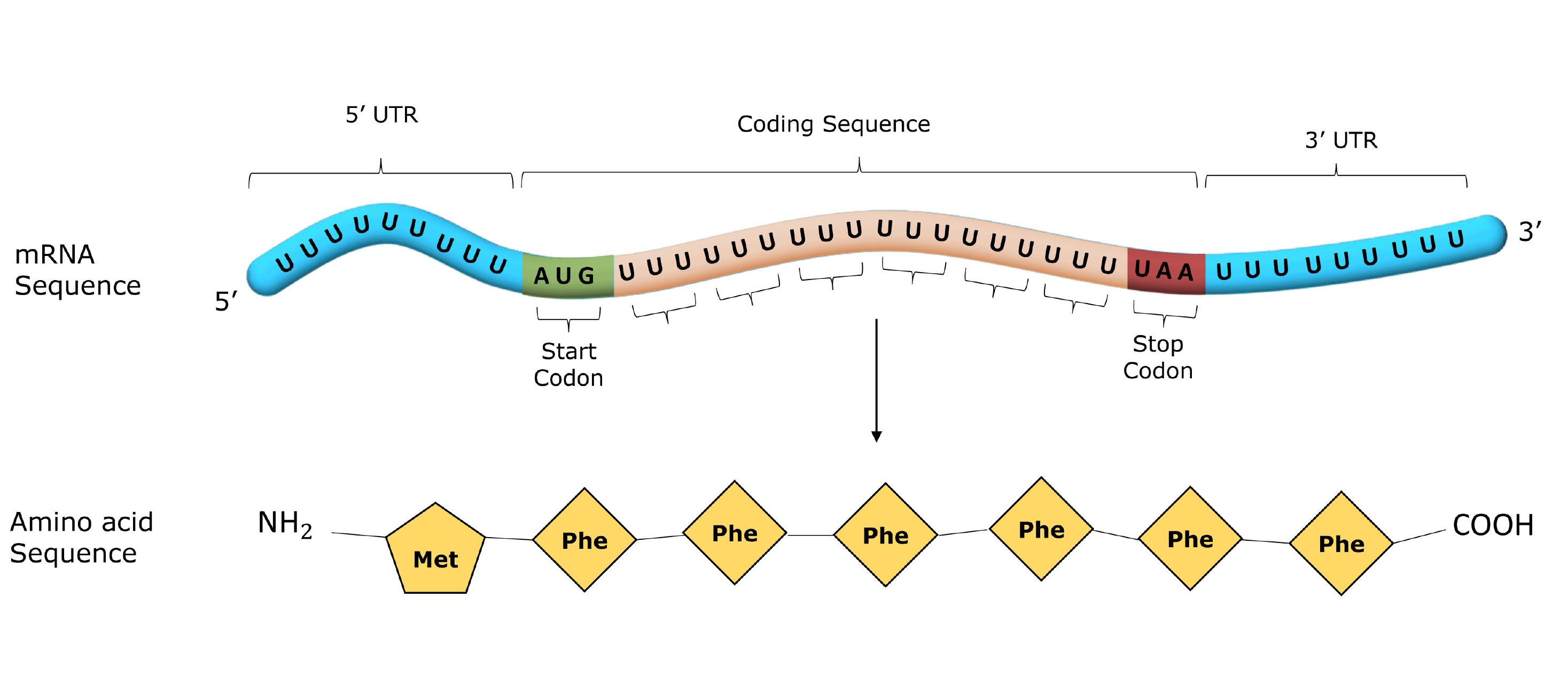}
\end{center}
\caption{A mRNA template with a homogeneous (poly-U) coding sequence and the corresponding sequence of amino acids (Phe) are shown to propose an experimental test of our theoretical predictions (see the text for details).}
\label{fig-polyU}
\end{figure}

We now propose an {\it in-vitro} experimental setup for testing our theoretical 
predictions reported in this paper. A sequence {\it homogeneous} mRNA template 
(for example, a poly-U, as shown schematically in Fig.\ref{fig-polyU}) would 
be ideally suited for this purpose. Such templates are used routinely for {\it 
in-vitro} experiments \cite{uemura10}. The coding sequence in such a mRNA 
template that is actually translated consists of $N_{c}$ number of identical 
codons UUU; in the poly-U template of Fig.\ref{fig-polyU}, $N_{c} = 6$. 
The coding sequence is preceded by a normal start codon (AUG) and is 
followed by a stop codon (UAA). The untranslated region (UTR) upstream 
from the start codon is required not only for assembling the ribosome from 
its subunits but also for stabilizing the pre-initiation complex. At the
3'-end, the stop codon is followed by a sequence of noncoding
codons UUU; this region merely ensures the absence of any `edge effect', 
i.e. the translation is not affected when the ribosome approaches the 
3'-end of the codon sequence. 

The optical method proposed here exploits labelling of the four species of 
tRNA molecules by four different fluorescent dyes of four distinct colors.
Each UUU codon codes for the amino acid phenylalanine(abbreviated Phe or F). 
Targeted (site specific) mutation at the editing site of the aatRNA 
synthetase would produce a defective variant of the enzyme whose editing 
mechanism has been disabled. The cognate tRNA molecules labelled by 
{\it red} fluorescent dye should be charged with amino acid Phe by the wild 
type aatRNA synthetase. In contrast, the cognate tRNA molecules labelled by 
{\it green} fluorescent dye should be separately charged with some amino 
acid other than Phe by the mutated aatRNA synthetase. Although the latter 
charging reaction is expected to proceed at a much slower rate than the 
former, yield can still be significant if the concentration of the amino 
acid substrate is sufficiently high. Thus, red and green fluorescence would 
signal correctly charged and mis-charged cognate tRNA species, respectively. 

A good choice for the corresponding near-cognate tRNA would be tRNALeu 
because it is cognate for the codon CUU which codes for leucine (abbreviated 
L). The correctly charged near-cognate tRNA can be labelled by orange dye 
while the non-cognate tRNA can be labelled by yellow dye. 
The color of the fluorescence pulse identifies the monomer species that 
elongates the polypeptide by one unit; monitoring the colors of the 
fluorescence pulses, one would get an estimate of mis-sense error 
arising separately from mis-charging of tRNA and mis-reading of mRNA . 
Moreover, the time interval between the arrival of the successive 
aa-tRNA molecules provides an estimate of the dwell times of the
ribosome.  Usually the coding sequence of such poly-U mRNA strands is 
quite short.  Therefore, for collecting enough data to extract the DDD, 
the experiment has to be repeated sufficiently large number of times.

\section{Summary and conclusion} 

In this paper we have developed a theoretical model that includes  
both the effects of mis-charging of tRNA and mis-reading of mRNA 
during the elongation cycle of gene translation. It also allows 
explicit distinction between (i) correctly charged cognate tRNA, 
(ii) incorrectly charged cognate tRNA, (iii) correctly charged 
near-cognate tRNA, and (iv) correctly charged non-cognate tRNA. 
For each of the four species, the master equations capture only 
the essential steps of the elongation cycle. From these equations 
we obtain the distribution of the dwell times of the ribosome at 
successive codons which is identical to the distribution of the times 
taken to incorporate the successive amino acids in the growing protein. 
From the steady-state solutions of the master equations we also derive
exact analytical formulae (\ref{eq-phiexp})-(\ref{eq-epmrexp}) 
that characterize various 
aspects of the erroneous protein polymerization process, particularly 
the average speed and fidelity of polymerization. The average 
speed of the ribosome, i.e., the average rate (\ref{eq-onebyV}) of 
elongation of the protein, is an interesting generalization of the 
Michaelis-Menten equation that governs the average rate of 
a very simple enzymatic reaction.
Some important implications of the analytical results reported here 
have been emphasized by plotting the results graphically. 
In particular, the plots show how with increasing rates of rejection 
of the near-cognate and non-cognate tRNAs the relative contribution 
of the mis-charged cognate tRNAs to the overall mis-sense error increases. 

Most of the experimental works on mis-reading error have been carried out 
for bacteria. So far as the eukaryotes are concerned, a comprehensive 
analysis of translational mis-reading error in budding yeast has been 
reported by Kramer et al.\cite{kramer10}. 
Mis-charging of tRNA and the failure of the editing mechanisms have been 
investigated separately for a long time \cite{ling09,yadavalli12}.  
However, to our knowledge, the relative contribution of mis-charging error 
to the overall mis-sense error has not been measured quantitatively in 
any experiment on translational kinetics. It is worth pointing out that a 
mis-charging error is not always detrimental for biological function of a 
cell and are believed to play some regulatory roles under special conditions 
\cite{pan13,pouplana14,ruan08,fan15}. 

The dependence of the frequencies of mis-reading error on the 
codon usage and tRNA concentration have been investigated 
extensively in the past 
\cite{garai09,fluitt07,zouridis08,basu07,shah10,rudorf15}; 
typical frequencies of mis-reading error can be as high as 
1 in $10^{3}$ \cite{kramer07}.
But, to our knowledge, mis-charging has not been incorporated so far in 
any mathematical model of kinetics of translation because under 
normal circumstances mis-charging error is as low as 1 in $10^5$ 
(or even lower). But, when subjected to various types of stress, 
at least ten fold increase in mis-charging has been observed 
\cite{netzer09}.  
The model and the analytical formulae derived here 
will be useful in analyzing the data collected in future experiments 
that might be performed for investigation of the same phenomenon.

Next we point out some features of the model that should be reflected 
in the experimental set up which may, in near future, be analyzed 
with the analytical formulae reported in this paper. For a natural 
mRNA molecule, because of its sequence inhomogeneity, the 
identity of the cognate tRNA keeps changing from one codon to 
another. On the other hand, the rate constants in our mathematical 
derivation is based on the assumption that the rates are independent 
of the position of the ribosome, i.e., independent of the identity of the 
codons. Thus, the sequence heterogeneity of natural mRNA molecules 
make those unsuitable for direct comparison with the analytical formulae 
reported here. Nevertheless, the model can be simply extended by 
assigning codon-dependent rate constants; but, in that case the results 
cannot be derived analytically (with closed form mathematical expressions) 
although all the quantities can be evaluated numerically. Since no 
experimental data for direct comparison is available at present, we have 
not carried out numerical study of the sequence-dependent model. 

As an alternative to sequence inhomogeneous real mRNA, a synthetic 
mRNA with homogeneous sequence can be used to directly test the 
validity of our {\it analytical formulae} reported here. For example, poly-U, 
along with the necessary start-, stop codons and untranslated region 
(UTR) upstream from the start codon \cite{sharma11a} could be a 
good candidate for this purpose. For any study of mis-sense error, 
the tRNA species which contribute the successive amino acids of the 
growing protein have to be identified. Fluorescence-based optical techniques 
\cite{uemura10,chen12} seem to be ideally suited for this purpose. 
The model will have to be extended in future also to incorporate the 
effects of microenvironment, cell cycle phase, etc. on translation. 
We hope that the relative quantitative contributions of mis-charging of 
tRNA and mis-reading of mRNA will be measured experimentally in 
near future and the analytical formulae derived here will find use 
in analyzing the experimental data. The experimental data will also 
guide extension of the model to make it more realistic by capturing 
features that are missing from the simple version reported in this 
paper.

\section*{Acknowledgements} DC thanks Joachim Frank, Ruben Gonzalez Jr. 
and Michael Ibba for useful correspondence. We also thank Joachim Frank, 
Adil Moghal and Alex Mogilner for valuable comments on a draft of this 
manuscript. 
We thank the anonymous referees for useful suggestions and for drawing our 
attention to a very recent paper.
This work is supported by ``Prof. S. Sampath Chair'' professorship (DC) 
and a J.C. Bose National Fellowship (DC). DC also acknowledges hospitality 
of the Biological Physics Group of the Max-Planck Institute for the Physics 
of Complex Systems at Dresden, under the Visitors Program, during the 
initial stages of this work. 
\clearpage
\begin{appendix}
\section{Master Equations and Steady-state probabilities} 
\label{sec-ssprobs}

The master equations governing the time evolution of the probabilities can be written as :
\begin{eqnarray}
\dfrac{dP_{1}(t)}{dt} & = & -(\omega_{a}+\omega^{'}_{a}+\omega^{''}_{a}+\omega^{'''}_{a})P_{1}(t)\nonumber \\
& + & \omega_{r1} P_{2}(t) + \omega^{'}_{r1}P^{'}_{2}(t) + \omega^{''}_{r1}P^{''}_{2}(t) \nonumber \\
& + & \omega^{'''}_{r1}P^{'''}_{2}(t) + \omega_{r2}P_{3}(t) + \omega^{'}_{r2}P^{'}_{3}(t) \nonumber \\
& + & \omega^{''}_{r2}P^{''}_{3}(t) + \omega^{'''}_{r2}P^{'''}_{3}(t) \nonumber \\
& + & \omega_{h2}P_{5}(t) + \Omega_{h2}P^{*}_{5}(t) \nonumber \\
\dfrac{dP_{2}(t)}{dt} &=& \omega_{a}P_{1}(t) - (\omega_{r1} + \omega_{h1})P_{2}(t) \nonumber \\
\dfrac{dP_{3}(t)}{dt} &=& \omega_{h1}P_{2}(t) - (\omega_{p} + \omega_{r2})P_{3}(t) \nonumber \\
\dfrac{dP_{4}(t)}{dt} &=& \omega_{p}P_{3}(t) + \omega^{'}_{p}P^{'}_{3}(t) + \omega_{br}P_{5}(t) -  \omega_{bf}P_{4}(t) \nonumber \\
\dfrac{dP_{5}(t)}{dt} &=& \omega_{bf}P_{4}(t) - (\omega_{h2} + \omega_{br})P_{5}(t) \nonumber \\
\dfrac{dP^{'}_{2}(t)}{dt} &=& \omega^{'}_{a}P_{1}(t) - (\omega^{'}_{r1} + \omega^{'}_{h1})P^{'}_{2}(t) \nonumber \\
\dfrac{dP^{'}_{3}(t)}{dt} &=& \omega^{'}_{h1}P^{'}_{2}(t) - (\omega^{'}_{r2} + \omega^{'}_{p})P^{'}_{3}(t) \nonumber \\
\dfrac{dP^{''}_{2}(t)}{dt} &=& \omega^{''}_{a}P_{1}(t) - (\omega^{''}_{r1} + \omega^{''}_{h1})P^{''}_{2}(t) \nonumber \\
\dfrac{dP^{''}_{3}(t)}{dt} &=& \omega^{''}_{h1}P^{''}_{2}(t) - (\Omega_{p} + \omega^{''}_{r2})P^{''}_{3}(t) \nonumber \\
\dfrac{dP^{'''}_{2}(t)}{dt} &=& \omega^{'''}_{a}P_{1}(t) - (\omega^{'''}_{h1} + \omega^{'''}_{r1})P^{'''}_{2}(t) \nonumber \\
\dfrac{dP^{'''}_{3}(t)}{dt} &=& \omega^{'''}_{h1}P^{'''}_{2}(t) - (\Omega^{'}_{p} + \omega^{'''}_{r2})P^{'''}_{3}(t) \nonumber \\
\dfrac{dP^{*}_{4}(t)}{dt} &=& \Omega_{p}P^{''}_{3}(t) + \Omega_{br} P^{*}_{5}(t) + \Omega^{'}_{p}P^{'''}_{3}(t) - \Omega_{bf}P^{*}_{4}(t) \nonumber \\
\dfrac{dP^{*}_{5}(t)}{dt} &=& \Omega_{bf}P^{*}_{4}(t) - (\Omega_{br} + \Omega_{h2})P^{*}_{5}(t) 
\label{eq-master}
\end{eqnarray}
with the normalization condition 
\begin{equation}
P_{1}(t)+P_{2}(t)+P_{3}(t)+P_{4}(t)+P_{5}(t)+P'_{2}(t)+P'_{3}(t)+P''_{2}(t)+P''_{3}(t)+P'''_{2}(t)+P'''_{3}(t)+P^{*}_{4}(t)+P^{*}_{5}(t)=1.
\end{equation}

The steady-state solutions $P_{\mu}$ of the equations (\ref{eq-master}) are given by
\begin{equation}
P_{1}=\dfrac{1}{X1}
\end{equation}
where
\begin{eqnarray*}
X1 &=& A \bigg [\dfrac{1}{\omega_{h1}}\bigg(1+\dfrac{\omega_{r2}}{\omega_p}\bigg)+\dfrac{1}{\omega_p}+\dfrac{1}{\omega_{bf}}\bigg(1+\dfrac{\omega_{br}}{\omega_{h2}}\bigg)+\dfrac{1}{\omega_{h2}} \bigg ] + \\
&& B \bigg [\dfrac{1}{\omega_{h1}'}\bigg(1+\dfrac{\omega_{r2}'}{\omega_p'}\bigg)+\dfrac{1}{\omega_p'}+\dfrac{1}{\omega_{bf}}\bigg(1+\dfrac{\omega_{br}}{\omega_{h2}}\bigg)+\dfrac{1}{\omega_{h2}} \bigg ] + \\
&& C \bigg [\dfrac{1}{\omega_{h1}''}\bigg(1+\dfrac{\omega_{r2}''}{\Omega_p}\bigg)+\dfrac{1}{\Omega_p}+\dfrac{1}{\Omega_{bf}}\bigg(1+\dfrac{\Omega_{br}}{\Omega_{h2}}\bigg)+\dfrac{1}{\Omega_{h2}} \bigg ] + \\
&& D \bigg [\dfrac{1}{\omega_{h1}'''}\bigg(1+\dfrac{\omega_{r2}'''}{\Omega_p'}\bigg)+\dfrac{1}{\Omega_p'}+\dfrac{1}{\Omega_{bf}}\bigg(1+\dfrac{\Omega_{br}}{\Omega_{h2}}\bigg)+\dfrac{1}{\Omega_{h2}} \bigg ]
\end{eqnarray*}

\begin{equation}
P_{2} = A \bigg [\dfrac{1}{\omega_{h1}}\bigg(1+\dfrac{\omega_{r2}}{\omega_p}\bigg) \bigg] P_{1}
\end{equation}

\begin{equation}
P_{3} = A \bigg [\dfrac{1}{\omega_{p}}\bigg] P_{1}
\end{equation}

\begin{equation}
P_{4} = (A+B) \bigg [\dfrac{1}{\omega_{bf}}\bigg(1+\dfrac{\omega_{br}}{\omega_{h2}}\bigg)\bigg] P_{1}
\end{equation}

\begin{equation}
P_{5} = (A+B) \bigg [\dfrac{1}{\omega_{h2}}\bigg] P_{1}
\end{equation}

\begin{equation}
P_{2}' = B \bigg [\dfrac{1}{\omega_{h1}'}\bigg(1+\dfrac{\omega_{r2}'}{\omega_p'}\bigg)\bigg] P_{1}
\end{equation}

\bigskip
\begin{equation}
P_{3}' = B \bigg [\dfrac{1}{\omega_{p}'}\bigg] P_{1}
\end{equation}

\begin{equation}
P_{2}'' = C \bigg [\dfrac{1}{\omega_{h1}''}\bigg(1+\dfrac{\omega_{r2}''}{\Omega_p}\bigg)\bigg] P_{1}
\end{equation}

\begin{equation}
P_{3}'' = C \bigg [\dfrac{1}{\Omega_{p}}\bigg] P_{1}
\end{equation}

\begin{equation}
P_{2}''' = D \bigg [\dfrac{1}{\omega_{h1}'''}\bigg(1+\dfrac{\omega_{r2}'''}{\Omega_p'}\bigg)\bigg] P_{1}
\end{equation}

\begin{equation}
P_{3}''' = D \bigg [\dfrac{1}{\Omega_{p}'}\bigg] P_{1}
\end{equation}

\begin{equation}
P^{*}_{4} = (C+D) \bigg [\dfrac{1}{\Omega_{bf}}\bigg(1+\dfrac{\Omega_{br}}{\Omega_{h2}}\bigg)\bigg] P_{1}
\end{equation}

\begin{equation}
P^{*}_{5} = (C+D) \bigg [\dfrac{1}{\Omega_{h2}}\bigg] P_{1}
\end{equation}


\clearpage

\section{Intuitive derivation of the expression for average rate of translation}
\label{sec-cleland}

Following Cleland's  approach \cite{cleland75} for replacing complex network of biochemical pathways 
by a simpler equivalent network and deriving the effective rates of the transitions of that network, we 
derive the expression for the average velocity of the ribosome in our model. 
To illustrate the method, we consider a simpler reaction
\begin{equation*}
X  \mathop{\rightleftharpoons}^{k_{1}}_{k_{-1}} Y \mathop{\rightarrow}^{k_{2}} Z
\end{equation*}
\hfill \break
The effective rate constant,  $k_{1}'$, for $X {\mathop{\rightarrow}}^{k_{1}'} Y$, is given by
\begin{equation*}
k_{1}'=\dfrac{k_{1}k_{2}}{k_{-1}+k_{2}}
\end{equation*}
The same treatment can be applied to our model. 

Let us first assume that only correctly charged cognate tRNAs are present (i.e. assuming that mischarged cognate, near cognate and non cognate tRNAs are absent in the surrounding). For the five consecutive steps of the cycle, we denote the transit times by $T_{1}$, $T_{2}$, $T_{3}$, $T_{4}$ and $T_{5}$, respectively. It is straightforward to see that

\begin{eqnarray}
T_{1}&=& \dfrac{1}{\omega_{a}}\bigg(1+\dfrac{\omega_{r1}}  {\omega_{h1}}\bigg)\bigg(1+\dfrac{\omega_{r2}}{\omega_p}\bigg) = \frac{1}{A} \\
T_{2}&=& \dfrac{1}{\omega_{h1}}\bigg(1+\dfrac{\omega_{r2}}{\omega_p}\bigg)\\
T_{3}&=& \dfrac{1}{\omega_{p}}\\
T_{4}&=& \dfrac{1}{\omega_{bf}}\bigg(1+\dfrac{\omega_{br}}{\omega_{h2}}\bigg)\\
T_{5}&=& \dfrac{1}{\omega_{h2}}\\
\end{eqnarray}
Therefore, for the above cycle, i.e., when only correctly charged cognate tRNA molecules are present in the surrounding, the average velocity of the ribosome would be
\begin{eqnarray}
\dfrac{1}{V_{c1}} &=& T_{c1}= T_{1}+T_{2}+T_{3}+T_{4}+T_{5}
\end{eqnarray} 
Similarly, for the other three cycles we can specify the transit times in a similar manner. For mischarged cognate tRNA, the corresponding symbols are $T'_{1},T'_{2},T'_{3},T'_{4},T'_{5}$ respectively, while for near cognate tRNA, ths symbols are $T''_{1},T''_{2},T''_{3},T''_{4},T''_{5}$ respectively,. For non cognate tRNA, we have $T'''_{1},T'''_{2},T'''_{3},T'''_{4},T'''_{5}$ respectively.

Next let us consider the general case when all the four different substrates are present in the 
surrounding; the kinetics of the system is described by full model shown in fig.\ref{fig-ribomodel}. 
The transit times are analogous to resistances in electrical circuits, which means that for a series of reaction, the transit times are additive and for parallel reaction pathways, the reciprocals of the transit times are additive. Hence, the average velocity for the complete model is
\begin{eqnarray}
\dfrac{1}{V_{tot}}=T_{tot}&=&\dfrac{A}{A+B+C+D}\Bigg (T_{1,eff}+T_{2}+T_{3}+T_{4}+T_{5} \Bigg )\\
&+&\dfrac{B}{A+B+C+D}\Bigg (T_{1,eff}+T'_{2}+T'_{3}+T'_{4}+T'_{5} \Bigg )\\
&+&\dfrac{C}{A+B+C+D}\Bigg (T_{1,eff}+T''_{2}+T''_{3}+T''_{4}+T''_{5} \Bigg )\\
&+&\dfrac{D}{A+B+C+D}\Bigg (T_{1,eff}+T'''_{2}+T'''_{3}+T'''_{4}+T'''_{5} \Bigg )
\end{eqnarray}
where 
\begin{eqnarray}
 \dfrac{1}{T_{1,eff}}= \dfrac{1}{T_{1}}+\dfrac{1}{T_{1}'}+\dfrac{1}{T_{1}''}+\dfrac{1}{T_{1}'''} \equiv A+B+C+D
\end{eqnarray}
\end{appendix}

\clearpage 
\section{Master Equations and derivation of dwell time distribution} 
\label{sec-dwell}

The master equations governing the time evolution of the probabilities are identical to those given in Appendix A, except for the following: 
\begin{eqnarray}
\dfrac{dP_{1}(t)}{dt} & = & -(\omega_{a}+\omega^{'}_{a}+\omega^{''}_{a}+\omega^{'''}_{a})P_{1}(t)\nonumber \\
& + & \omega_{r1} P_{2}(t) + \omega^{'}_{r1}P^{'}_{2}(t) + \omega^{''}_{r1}P^{''}_{2}(t) \nonumber \\
& + & \omega^{'''}_{r1}P^{'''}_{2}(t) + \omega_{r2}P_{3}(t) + \omega^{'}_{r2}P^{'}_{3}(t) \nonumber \\
& + & \omega^{''}_{r2}P^{''}_{3}(t) + \omega^{'''}_{r2}P^{'''}_{3}(t) \nonumber \\
\dfrac{d\widetilde{P_{1}}(t)}{dt} & = & \omega_{h2}P_{5}(t) + \Omega_{h2}P^{*}_{5}(t)
\label{eq-masterDWELL}
\end{eqnarray}
and the normalization condition becomes 
\begin{equation}
P_{1}(t)+P_{2}(t)+P_{3}(t)+P_{4}(t)+P_{5}(t)+P'_{2}(t)+P'_{3}(t)+P''_{2}(t)+P''_{3}(t)+P'''_{2}(t)+P'''_{3}(t)+P^{*}_{4}(t)+P^{*}_{5}(t)+\widetilde{P_{1}}(t)=1.
\end{equation}
Here, we take the initial conditions to be $P_{1}(0)=1$, and $P_{2}(0)=P_{3}(0)=P_{4}(0)=P_{5}(0)=P'_{2}(0)=P'_{3}(0)=P''_{2}(0)=P''_{3}(0)=P'''_{2}(0)=P'''_{3}(0)=P^{*}_{4}(0)=P^{*}_{5}(0)= \widetilde{P_{1}}(0)=0$.

\section{References}


\end{document}